\documentstyle[12pt]{article}

\begin{document}

\title{Geometric Models of the  Quantum Relativistic Rotating Oscillator}

\author{Ion I. Cot\u aescu\\ The West University of Timi\c soara,\\
        V. P\^ arvan Ave. 4, RO-1900 Timi\c soara, Romania}

\date{\today}

\maketitle

\begin{abstract}
A family of geometric models of quantum relativistic rotating oscillator is 
defined by using a set of one-parameter deformations of the static (3+1) 
de Sitter or anti-de Sitter metrics.  
It is shown that all these models lead to the usual isotropic harmonic 
oscillator in the non-relativistic limit, even though their relativistic 
behavior is  different. As in the case of the (1+1) models \cite{COTA}, these 
will have even countable energy spectra or mixed ones, with a  finite discrete
sequence and a continuous part. In addition, all these spectra, 
except that of the pure anti-de Sitter model, will have a fine-structure, 
given by a rotator-like term.

%Pacs 04.62.+v, 03.65.Ge
\end{abstract}
\
\

In the general relativity  the relativistic  (three-dimensional isotropic) 
harmonic oscillator (RHO) is defined as a free system on the  
anti-de Sitter  static background \cite{P1,P2,A1,A2,A3}. There exists a metric 
\cite{P2} which 
reproduces the classical equation of motion of the non-relativistic (isotropic) 
harmonic oscillator (NRHO). Moreover, the corresponding quantum system 
represented by a free scalar field on this background has an
equidistant energy spectrum with a ground state energy larger 
than, but approaching $3\omega/2$ in the non-relativistic limit (in natural 
units, $\hbar =c =1$) \cite{M}. Thus, the static anti-de Sitter classical and 
quantum geometric
models reproduce all the properties of the NRHO giving as unique relativistic 
effect the value of the ground state energy of the quantum RHO. For this reason 
it is interesting to look for new geometric models which should behave as the 
NRHO in the non-relativistic limit but having strong specific  relativistic 
effects. In a previous article \cite{COTA} we have shown  that such 
models can be constructed in (1+1) dimensions by 
using appropriate one-parameter conformal transformations of the anti-de Sitter 
or de Sitter  static metrics.  Thus, we 
have found  a set of (1+1) models with countable energy spectra, and another 
one having only a finite number of energy levels and a continuous 
spectrum. It remains us to see if these properties could be recovered also in 
the case of the (3+1) models.

This article is devoted to this problem. In (3+1) dimensions the 
 conformal transformations will not suffice to give  natural generalizations 
of our (1+1) metrics \cite{COTA}. Therefore, we are forced to use suitable 
{\it deformations} of the de Sitter or  anti-de Sitter  metrics. Let us start 
with the anti-de Sitter static metric proposed in Ref. \cite{P2}   
\begin{equation}\label{(m1)}
ds^{2}=\frac{1}{1-\omega^{2} r^{2}}dt^{2}-\frac{1}{1-\omega^{2}
r^{2}}\left( \delta_{ij}+ \frac{\omega^{2} x^{i}x^{j}}{1-\omega^{2}r^{2}}
\right)dx^{i}dx^{j},
\end{equation}
where $x^{i}$ are the Cartesian space coordinates ($i,j,..=1,2,3$) and 
$r=\sqrt{x^{i}x^{i}}$. This will be deformed into a family of metrics 
depending on a real parameter, $\lambda$, as follows  
\begin{equation}\label{(m)}
ds^{2}=g_{00}dt^{2}+g_{ij}dx^{i}dx^{j}=
\frac{\alpha}{\beta}dt^{2}-  
\frac{1}{\beta}\left(\delta_{ij}+
\frac{\omega^{2}}{\beta}x^{i}x^{j}\right) dx^{i}dx^{j},
\end{equation}
where 
\begin{equation}
\alpha=1+(1+\lambda)\omega^{2}r^{2}, \qquad 
\beta=1+\lambda\omega^{2}r^{2} 
\end{equation}
This family contains  deformed de Sitter metrics for $\lambda>0$,  
deformed anti-de Sitter metrics for $\lambda<0$ and a deformed Minkowski 
metric corresponding to $\lambda=0$. The  exact anti-de Sitter metric (\ref{(m1)}) can 
also be obtained for  $\lambda = -1$. The deformed anti-de Sitter metrics are 
singular for $r=r_{0}=1/\omega \sqrt{-\lambda}$ which is just the radius of the 
event horizon  for an observer situated at $x^{i}=0$. In the case of 
non-negative $\lambda$ this is $r_{0}=\infty$. In the following, we shall 
consider the free motion on the domain $D$ of the space coordinates bounded 
by the event horizon, i.e. $r\in  [0,r_{0})$. Here the equations of the 
classical free motion given by these metrics will be similar to that of 
the NRHO up to  a Coriolis-like term (linear in $\dot x^{i}$)  which will 
contribute for all $\lambda \not= -1$. This  produces a specific  
rotation  which  dissapears in the non-relativistic limit. For this reason,  
all the systems with $\lambda \not= -1$ will be referred as   
relativistic rotating oscillators (RRO), understanding that the RHO is only that 
of the  anti-de Sitter metric. We note that all these metrics are invariant 
under time  translations and  space rotations. Consequently, the energy and 
the angular momentum are conserved.

Our aim is to investigate the  properties of the quantum free motions of a 
spinless particle of mass $M$ on the backgrounds given by (\ref{(m)}) and to 
show that all these models lead to the quantum NRHO in the non-relativistic 
limit.  Let us consider the scalar 
field  $\phi$, defined on $D$, minimally coupled with the gravitational field 
\cite{B1}. Because of the energy conservation, the Klein-Gordon equation
\begin{equation}\label{(kg)}
\frac{1}{\sqrt{-g}}\partial_{\mu}\left(\sqrt{-g}g^{\mu\nu}\partial_{\nu}\phi
\right) + M^{2}\phi=0
\end{equation}
where $g=\det(g_{\mu\nu})$, admits a set of fundamental solutions (of positive 
and negative frequency) of the form
\begin{equation}\label{(sol)}
\phi_{E}^{(+)}=\frac{1}{\sqrt{2E}}e^{-iEt}U_{E}(x), \quad 
\phi^{(-)}=(\phi^{(+)})^{*},
\end{equation}
which must be orthogonal with respect to the relativistic scalar product 
\cite{B1} 
\begin{equation}\label{(sp1)}
<\phi,\phi'>=i\int_{D}d^{3}x\sqrt{-g}g^{00}\phi^{*}\stackrel{\leftrightarrow}{\partial_{0}} \phi'.
\end{equation}
We shall start with the metric (\ref{(m)}) in spherical coordinates
\begin{equation}\label{(ms)}
ds^{2}=\frac{\alpha}{\beta}dt^{2} -
\frac{\alpha}{\beta^{2}}dr^{2} -
\frac{r^{2}}{\beta}(d\theta^{2}+\sin^{2}\theta d\phi^{2})
\end{equation}
assuming that  
\begin{equation}
U(x)=R_{E,l}(r)Y_{lm}(\theta, \phi),
\end{equation}
where $Y_{lm}$ are the usual spherical harmonics and $R_{E,l}$ are the radial 
wave functions. The conservation of the angular momentum allows to separate the 
variables of the Klein-Gordon equation obtaining the radial equation
\begin{equation}\label{(kg1)}
\beta\frac{d^{2}R}{dr^{2}}+\frac{\beta}{r}\left(1+ \frac{1}{\beta}
\right) \frac{dR}{dr}-
\frac{\alpha}{r^{2}}l(l+1)R+\left(E^{2}-\frac{\alpha}{\beta} M^{2}\right)R=0
\end{equation}
Moreover, from (\ref{(sp1)}) it results that the scalar product of the 
radial wave functions is
\begin{equation}\label{(psc)}
<R,R'>=\int_{0}^{r_{0}}\frac{r^{2}dr}{(1+\lambda\omega^{2}r^{2})^{3/2}}R^{*}R'.
\end{equation}
In the following we shall  derive the energy spectrum and  the wave functions 
up to normalization factors.

When $\lambda=0$ the equation (\ref{(kg1)}) becomes
\begin{equation}
\left(-\frac{d^{2}}{dr^{2}} - \frac{2}{r}\frac{d}{dr}+\frac{l(l+1)}{r^{2}}
+M^{2}\omega^{2}r^{2}\right)R^{0}_{E,l}=(E^{2}-M^{2}-\omega^{2}l(l+1))
R^{0}_{E,l},
\end{equation}
from which it results that, in this case, the radial wave functions coincide 
with those of the NRHO,  having the form 
\begin{equation}\label{(u0)}
R^{0}_{n_{r},l}=N^{0}_{n_{r},l}r^{l}F(-n_{r},l+\frac{3}{2}, M\omega r^{2})
e^{-M\omega r^{2}/2}
\end{equation} 
where $F$ is the confluent hypergeometric function and $n_{r}=0,1,2,...$ 
is the radial 
quantum number. Now, by introducing the main quantum number, $n=2n_{r}+l$,  
we see that the energy levels 
\begin{equation}\label{(s1)}
{E_{n,l}}^{2}=M^{2}+2M\omega(n+\frac{3}{2})+\omega^{2}l(l+1), 
\end{equation} 
depend on $n=0,1,2,...$ and on the angular momentum quantum number,$l$, which 
appears in a rotator-like term. It is obvious that $l$ will take the even 
values from $0$ to $n$ if $n$ is even and the odd values from $1$ to $n$ for 
each odd $n$. Thus, the energies remain degenerated only upon the quantum 
number $m$. We note that the rotator-like term is of the order $1/c^{2}$ and, 
consequently, this will not contribute in the non-relativistic limit, when 
$\omega/M \rightarrow 0$.
Moreover, one can verify that, in this limit, the spectrum given by  (\ref{(s1)}) 
becomes just the NRHO traditional one. 

In the general case of any $\lambda\not=0$ it is convenient to use the new 
variable $y=-\lambda\omega^{2}r^{2}$, and the notations
\begin{equation}\label{(nu)}
\epsilon= \frac{E}{\lambda\omega}, \quad \mu=\frac{M}{\lambda\omega}, \quad 
\nu=\frac{1}{4}\left[(1+\lambda)\mu^{2}-\lambda \epsilon^{2} +
\frac{1+\lambda}{\lambda}l(l+1)\right].
\end{equation}
We shall look for a solution of the form
\begin{equation}\label{(100)}
R(y)=N(1-y)^{p}y^{s}F(y),
\end{equation}
where $p$ and $s$ are real numbers and $N$ is the normalization factor. After 
a few manipulation we find that, for  
\begin{equation}\label{(pau)}
2s(2s+1)=l(l+1), \qquad 4p^{2}-6p-\mu^{2}=0,
\end{equation}
the equation (\ref{(kg1)}) transforms into the following hypergeometric 
equation:
\begin{equation}
y(1-y)F_{,yy}+[2s+\frac{3}{2}-y(2p+2s+1)]F_{,y}-[(p+s)^{2}-\nu]F=0.
\end{equation}
This has the solution \cite{B2}
\begin{equation}\label{(F1)}
F=F(p+s-\sqrt{\nu}, p+s+\sqrt{\nu}, 2s+\frac{3}{2}, y),
\end{equation}
which depends on the possible values of the parameters $p$ and $s$. From 
(\ref{(pau)}) it follows that  
\begin{equation}
s=\frac{l}{2}, \qquad p=p_{\pm}=\frac{3}{4}\left[ 1\pm \sqrt{1+ 
\left(\frac{2\mu}{3}\right)^{2}}\right].
\end{equation}
since the other possible solution of the first equation of (\ref{(pau)}), 
$s=-(l+1)/2$, produces singularities at $r=0$. Furthermore, 
we observe that for 
\begin{equation}\label{(quant1)}
\nu=(p+s+n_{r})^{2}, \quad n_{r}=0,1,2...,
\end{equation}
$F$ reduces to a polynomial of degree $n_{r}$ in $y$.  
According to  these results, we can establish the general form of the 
solutions of (\ref{(kg1)}), which could be square integrable with respect to 
(\ref{(psc)}), namely 
\begin{equation}\label{(u1)}
R_{n_{r},l}(x)=N_{n_{r},l}(1+\lambda\omega^{2}r^{2})^{p}r^{l}F(-n_{r},
2p+l+n_{r}, l+\frac{3}{2}, -\lambda\omega^{2}r^{2}).
\end{equation}
By using  the quantum number $n=2n_{r}+l$ and the relations  
(\ref{(quant1)}), (\ref{(pau)}) and (\ref{(nu)}) we can derive the 
following formula of the energy levels:
\begin{equation}\label{(el)}
{E_{n,l}}^{2}=M^{2}-\lambda\omega^{2}[4p(n+\frac{3}{2})+n^{2}] +
(1+\lambda)\omega^{2}l(l+1). 
\end{equation}  
These are depending on $n$ and $l$ with the same selection rules as mentioned 
above in the case of $\lambda=0$. Herein we see that the rotator-like term 
of (\ref{(el)}),  which is of the order $1/c^{2}$,   
does contribute for all the models with $\lambda \not=-1$, 
vanishing only for the anti-de Sitter RHO. Now, it remains  to 
fix the  suitable values of $p$ for which $<R_{n_{r},l},R_{n_{r},l}> < \infty$, and 
to analyze the structure of the obtained spectra. 

Let us first take  $\lambda>0$. In this case $r_{0}=\infty$, and the 
solution (\ref{(u1)}) will be  square integrable only if $p=p_{-}$ and 
$n<-2p_{-}$. This means that the discrete spectrum is {\it finite}, 
with $n=0,1,2...n_{max}$, 
where $n_{max}$ is the integer part of  $3(\sqrt{1+4\mu^{2}/9} - 1)/2$. 
For $n=l=n_{max}$ we obtain the maximal energy level, $E_{max}$, which satisfies
\begin{equation}
{E_{max}}^{2}\le M^{2}\left(1+\frac{1}{\lambda}\right)^{2}+4p_{-}\omega^{2}
(1+\lambda) 
\end{equation}
We note that the number of the discrete levels is strongly dependent upon the 
value of $M/\lambda \omega$. For example, in the ultra-relativistic domain 
$\lambda \omega >M/2$ the value of $-2p_{-}$ is smaller than $1$ and, 
consequently, the discrete spectrum  will contain  only  the ground level with 
 $n=l=0$. On the other hand, from (\ref{(nu)}) it results that  for any $E$ 
and $l$  satisfying 
\begin{equation}\label{(spec)}
E^{2}-(1+\lambda)\omega^{2}l(l+1)\ge 
(1+\frac{1}{\lambda})M^{2}
\end{equation}
$\nu$ is  negative or zero and the hypergeometric functions (\ref{(F1)}) cannot be reduced 
to polynomials, but remain analytic for negative arguments. Therefore, the 
functions 
\begin{equation}
R_{\nu,l}=N_{\nu,s}(1+\lambda\omega^{2}r^{2})^{p_{-}}r^{l}F(p_{-}+
\frac{l}{2}-\sqrt{\nu},
p_{-}+\frac{l}{2}+\sqrt{\nu},l+\frac{3}{2},-\lambda\omega^{2}r^{2})
\end{equation} 
can be interpreted as the non-square integrable solutions corresponding to 
the {\it continuous} energy spectrum. According to (\ref{(spec)}) this is 
$[M \sqrt{1+1/\lambda}, \infty)$. What is interesting here is that, for the 
values of the parameters giving $n_{max}>0$, the highest levels of the discrete 
spectrum will overlap the continuous spectrum. In these conditions  
{\it spontaneous} transitions from the discrete spectrum into the continuous 
one become possible.

In the case of $\lambda<0$ the radial domain is finite the particle being 
confined to the spherical cavity of the radius $r_{0}=1/\omega \sqrt{-\lambda}$.  
The polynomial solutions (\ref{(u1)}) will be
square integrable over $[0,r_{0})$ only if they are regular at $r_{0}$. 
This require the choice  $p=p_{+}$  for which there are no restrictions on 
the range of $n$. Therefore, the discrete spectrum is countable. Moreover, in 
this case we have no continuous spectrum since the hypergeometric functions 
(\ref{(F1)}) generally diverge for $y\rightarrow 1$ (when $r\rightarrow r_{0}$).    
For the RHO ($\lambda=-1$) we obtain the same 
result as in Ref. \cite{M}, i.e.  the equidistant levels 
\begin{equation}
E_{n}=\omega(2p_{+}+n),\quad n=0,1,2,...
\end{equation} 
depending only on $n$.

All these results are similar with those obtained for the (1+1) models with 
conformal transformations of de Sitter or anti-de Sitter metrics \cite{COTA}. 
Moreover, we can 
prove that  the solutions we have obtained here are continuous with respect to 
$\lambda$, as in the case of the (1+1) models.
Indeed, when $\lambda \rightarrow 0$ then $p\sim-M/2\lambda \omega$ (since 
we have chosen $p=p_{-}<0$ for $\lambda>0$ and $p=p_{+}>0$ for $\lambda<0$).
Therefore, $n_{max}\sim M/\lambda\omega \rightarrow \infty$ which means that 
the finite discrete spectra of the models with $\lambda>0$ become countable 
while the continuous spectra disappear. Hence it is obvious that the discrete 
spectra given by (\ref{(el)}) become (\ref{(s1)}) when $\lambda \rightarrow 0$.
 Furthermore, we can convince ourselves that, in this limit, the radial wave functions 
(\ref{(u1)}) take the form (\ref{(u0)}) which coincides to that of the NRHO 
radial wave functions. Thus, it results that our solutions are continuous in  
$\lambda$. Based on this property we can calculate the 
non-relativistic limit  like in the (1+1) case, taking $\lambda \rightarrow 0$ 
and, in addition, $M\gg \omega$.  The conclusion is that all the models of 
RRO we have studied here have the same non-relativistic limit, namely the NRHO.

Finally we note that the RRO with $\lambda<0$ could be 
used as  models of geometric confinement. The advantage here is that we have 
the new arbitrary parameter $\lambda$ which allows us to choose the desired 
values of the frequency and of the radius of the spherical cavity. Moreover, 
the  discrete spectra of the RRO  have a fine-structure 
which could be useful in building new  models of this kind.

\end{document}